# Large Spontaneous Nonreciprocal Charge Transport in a Zero-Magnetization Antiferromagnet

Kenta Sudo[1, †, *], Yuki Yanagi[2], Mitsuru Akaki[1], Hiroshi Tanida[2], and Motoi Kimata[1, 3, **]

[1]*Institute for Materials Research, Tohoku University, Sendai, Miyagi 980-8577, Japan*
[2]*Liberal Arts and Sciences, Toyama Prefectural University, Imizu, Toyama 939-0398, Japan*
[3]*Advanced Science Research Center, Japan Atomic Energy Agency, Tokai, Ibaraki 319-1195, Japan*

[†]*Present address: Institute for Solid State Physics, University of Tokyo, Kashiwa 277-8581, Japan*

*ksudo@issp.u-tokyo.ac.jp, **kimata.motoi@jaea.go.jp

**Abstract**
Spontaneous breaking of time-reversal and spatial-inversion symmetries in solids triggers diverse intriguing phenomena. Although these phenomena have been extensively studied in insulators, similar investigations for metals remain limited. Herein, we report the observation and properties of spontaneous (i.e., zero-magnetic field) nonreciprocal charge transport in the zigzag intermetallic compound $NdRu_2Al_{10}$. This effect is attributed to the antiferromagnetic (AF) order, which can be interpreted as a magnetic toroidal dipole order. Our results reveal an excessively large nonreciprocal coefficient for this material, attributed to the strong effective magnetic field generated through *c–f* exchange interactions. The results also suggest that the nonreciprocal response of this material depends on the spin configurations of the AF domains. Overall, our findings are distinct from those previously reported for field-induced nonreciprocal charge transport and contribute to a comprehensive understanding of cross-correlations in symmetry-broken metals.

**Main text**
Spontaneous breaking of time-reversal symmetry (TRS) and spatial-inversion symmetry (SIS) leads to various intriguing phenomena in condensed matter physics. One well-known example is the magnetoelectric effect in multiferroics, where cross-correlations emerge between magnetism and dielectric polarization due to spontaneously broken symmetries [1,2]. However, in metals, dielectric polarization is screened by conduction electrons, leading to cross-correlations that differ from those observed in multiferroics. Notably, representative examples of cross-correlations observed in symmetry-broken metals include the recently discovered phenomena of current-induced magnetization [3–6] and nonreciprocal charge transport [7–19]. Among them, current-induced magnetization is a phenomenon in which a spin polarization is produced under an applied electrical current, while nonreciprocal charge transport deviates from Ohm's law and exhibits rectification effects even within a single metallic material without an interface. Based on the current understanding, these responses originate from asymmetric band structures in the momentum space, which arise from the simultaneous breaking of TRS and SIS. This condition has been artificially produced by applying an external magnetic field to a material with broken SIS, where TRS is exogenously broken by the applied magnetic field [7–19]. Drawing parallels with multiferroics, the spontaneous breaking of both TRS and SIS in metals can yield even richer physical properties without requiring a magnetic field. However, experimental investigations into these spontaneous symmetry-breaking effects in metals remain limited.

Nonreciprocal charge transport—a typical cross-correlation observed in symmetry-broken metals—is phenomenologically expressed by expanding Ohm's law, which involves replacing electrical resistivity $\rho$ by $\rho + \gamma j$, where the additional term $\gamma j$ depends on the current. The resulting expression is

$$E = (\rho + \gamma j)j = \rho j + \gamma j^2 = E^{\text{Ohm}} + E^{\text{NR}}, \qquad (1)$$

where $E$ [V/m], $\rho$ [Ωm], and $j$ [A/m²] denote the electric field, Ohmic electrical resistivity, and current density, respectively. Meanwhile, $\gamma$ [ΩA$^{-1}$m$^3$] represents the nonreciprocal resistivity. Once TRS and SIS are broken, $\gamma$ becomes finite and the resistivity depends on current polarity. In previous studies [7–19], nonreciprocal resistivity $\gamma$ was observed only in the presence of an external magnetic field $B$; consequently, the nonreciprocal term $E^{\text{NR}}$ was expressed as $\gamma j^2 B$. Conversely, the nonreciprocal charge transport reported herein originates from antiferromagnetic (AF) order in the absence of an external magnetic field (i.e., under zero-field condition) and residual magnetization. Therefore, we use $\gamma j^2$ instead of $\gamma j^2 B$ in Eq. (1) to describe zero-field nonreciprocal resistivity. This study further reveals that zero-field nonreciprocal resistivity is governed by the spin configuration of AF domains. These findings suggest the potential application of nonreciprocal transport in the electrical readout of AF domains, complementing previously reported optical detection methods [20–23].

Figure 1(a) illustrates the zigzag chain of NdRu$_2$Al$_{10}$ in the paramagnetic state, the target material of this study, with the space group *Cmcm* (No. 63)[24]. In this structure, Nd atoms coordinated by four Ru atoms form a zigzag structure. A key feature is the sublattice degrees of freedom, labeled as sublattice 1 and 2 in Fig. 1(a), arising from the glide symmetry along the *b*-axis. Consequently, the local SIS is broken at the Nd site (4$c$ site, $C_{2v}$), leading to an electric-field gradient direction that is opposite between the sublattices, as indicated by orange and blue arrows in Fig. 1(a). When assuming a large hopping integral between Nd sites within the same sublattice (dashed lines in Fig. 1(a)), each sublattice chain forms an independent Fermi surface (FS) [25]. Due to the electric-field gradient along the $b$-axis, the FS is expected to exhibit Rashba-type spin polarization confined to the $ac$-plane, as shown on the left side (labeled as "Para.") of Fig. 1(b). The spin polarization of the Fermi surface is solely determined by the local point group symmetry ($C_{2v}$) of Nd sites and is therefore not sensitive to the detail of magnetic structure[26]. Importantly, the sign of spin polarization is inverted between sublattices because of the staggered direction of electric-field gradient [25,27,28]. Since the global SIS is preserved by the $Cmcm$ ($D_{2h}$) symmetry in the paramagnetic state, the overall spin polarization cancels out, preventing nonreciprocal charge transport under a uniform magnetic field. However, when a sublattice-dependent effective magnetic field ($B_{eff}$) is applied oppositely to sublattices 1 and 2 (right side of Fig. 1(b)), the FSs shift in the same direction, producing nonreciprocal charge transport. This can be achieved by an AF order with an easy axis within the $ac$-plane. Specifically, for NdRu$_2$Al$_{10}$, since the AF easy axis is the $a$-axis, nonreciprocal transport is expected for current $I \parallel c$. This kind of nonreciprocal charge transport arising from a shit of Fermi surface was proposed in the previous work [7], and our description here can be regarded as its extension to systems with local inversion symmetry breaking.

On the other hand, the emergence of nonreciprocal charge transport due to AF order in the zigzag structure aligns with a recent theory based on multipole [27,29]. In NdRu$_2$Al$_{10}$, the AF state is characterized by a magnetic toroidal dipole, that breaks TRS and global SIS, along the zigzag chain, and thus nonreciprocal transport is anticipated along the toroidal moment ($c$-direction) [30] (See Supplemental Material at [URL will be inserted by publisher] for relationship between magnetic toroidal moment and nonreciprocal charge transport in NdRu$_2$Al$_{10}$ (see also references [19,27,29] therein)).

Figure 1(c) presents a schematic of our experimental setup. The scanning electron microscopy (SEM) image of a microfabricated device is shown in Fig. 1(c). [30] (See Supplemental Material at [URL will be inserted by publisher] for experimental method (see also references [7,24,31,32] therein). The main panel of Fig. 1(d) illustrates the temperature dependence of resistivity $\rho$ below room temperature. Metallic behavior, along with a decrease in resistivity, is observed in the whole temperature range, and a kink is observed around the Neel temperature, $T_N$ = 2.4 K [24,33] (inset of Fig. 1(d)). These results are consistent with the behavior observed in bulk crystals, indicating that the damage of the microfabrication process on crystal quality is minimal: both the electrical and magnetic properties of the bulk crystal are well-preserved in the fabricated devices.

Figure 1(e) shows the temperature dependence of the nonreciprocal resistivity ($\gamma$) from 1.5 to 4 K under zero magnetic field. In the paramagnetic state ($T > T_N$), where both TRS and SIS are preserved, $\gamma$ remains negligible. In the AF state ($T < T_N$), $\gamma$ gradually increases with decreasing temperature. As shown by the differently colored data in Fig. 1(e), $\gamma$ is independent of current density. This behavior is consistent with previous reports on nonreciprocal resistivity [7–13,15–19]. These results indicate that the observed nonreciprocal charge transport intrinsically arises from AF order in NdRu$_2$Al$_{10}$. Furthermore, magnetization measurements confirm the absence of spontaneous magnetization at $B = 0$ [24].

The theoretical model [27] predicts a divergent enhancement of the nonreciprocal response just below $T_N$, in contrast to the gradual development observed the experiment. This discrepancy is arising from the different Fermi-level positions: in the model calculation, the Fermi energy lies close to the band edge, so that the band shift along the vertical (energy) direction due to the antiferromagnetic order causes the band to cross the Fermi level, leading to a sharp change in nonreciprocal response. In the present metallic system, however, the Fermi level is much deeper, and such a band-crossing effect is likely negligible.

The dependence of nonreciprocal resistivity on AF domains provides further evidence confirming the zero-field nonreciprocal response originating from the AF order. Given that the switching between two AF domains with opposite spin configurations—i.e., ↑↓↑↓ (domain A) and ↓↑↓↑ (domain B)—is equivalent to a time-reversal operation, the nonreciprocal signals originating from domains A and B are anticipated to exhibit opposite-sign $\gamma$. To investigate this, we fabricated device #2 (Fig. 2(a)). This device was designed to have a considerably longer channel length compared to device #1 for examining the expected AF-domain-dependent behavior of $\gamma$. Figure 2(b) illustrates the temperature dependence of $\gamma$ for different voltage-contact pairs ($V_1$–$V_6$) indicated in Fig. 2(a). As shown in Fig. 2(b), the values of $\gamma$ for $V_1$ and $V_2$ are positive ($\gamma > 0$), while for $V_6$, $\gamma$ is negative ($\gamma < 0$). Furthermore, small positive values of $\gamma$ are observed for $V_3$ and $V_4$, while a small negative $\gamma$ value is observed for $V_5$. The similar absolute values of $\gamma$ for $V_1$, $V_2$, and $V_6$ (~ 3 × 10$^{-19}$ ΩA$^{-1}$m$^3$) suggest that these $\gamma$ values originate from AF domains with a single spin configuration, either ↑↓↑↓ or ↓↑↓↑, rather than from a mixture of these domains. In contrast, the small values of $\gamma$ observed in the intermediate voltage-contact pairs ($V_3$–$V_5$) indicate partially compensated signals owing to domain mixing. Based on these observations, we can estimate the mixing ratio between the A and B domains. Figure 2(c) presents the estimated domain ratio between domains A and B, indicating a gradual variation in the AF domains over a length scale of a few tens of micrometers. This length scale is reasonably consistent with previously reported AF domain sizes in other materials [22,23]. A schematic of the expected domain distribution is presented in Fig. 2(d).

We examined the magnetic-field response of $\gamma$ in each domain, as shown in the lower three panels of Fig. 2(e). These panels display $\gamma$ in domain A ($V_A$), domain B ($V_B$), and at the boundary ($V_{bound}$) (Fig. 2(d)) under magnetic fields applied along the $a$-axis (left), $b$-axis (middle), and $c$-axis (right). For comparison, the upper panels show the normalized Ohmic resistance $\rho/\rho_{B=0}$. While $\rho/\rho_{B=0}$ exhibits nearly identical field dependence across all domains, $\gamma$ shows distinct behavior. Specifically, $\gamma$ in $V_A$ and $V_B$ displays similar trends but with opposite signs (orange and blue lines, respectively), while $\gamma$ at $V_{bound}$ remains close to zero. The symmetric field dependence of $\gamma$ with respect to field reversal indicates that $\gamma$ behaves as an even function, unlike the odd-function behavior of previously reported field-induced nonreciprocal charge transport [7–19]. The characteristic field dependence of $\gamma$ correlates with magnetization reorientation under varying magnetic fields. When fields are applied along the $b$- or $c$-axes, $|\gamma|$ decreases and nearly vanishes above the saturation field $B_{sat}$, where Nd moments become uniformly polarized, eliminating the AF spin configuration ($B_{sat} \approx 7.2$ T for $B \parallel b$ and $B_{sat} \approx 5.2$ T for $B \parallel c$ at $T = 1.4$ K) [24]. This behavior suggests restoration of SIS in the saturation region. In contrast, for $B \parallel a$ (the easy axis), a metamagnetic transition at $B_M$ ($B_M \approx 2.6$ T at $T = 1.4$ K) [24] leads to a different $\gamma$ response. Here, $\gamma$ temporarily vanishes at $B_M$ and re-emerges, maintaining a nonzero value beyond $B_{sat}$ ($B_{sat} \approx 4.6$ T for $B \parallel a$ at $T = 1.4$ K) [24], as shown in the lower left panel of Fig. 2(e).

We investigate the origin of the characteristic behavior of $\gamma$ around $B_M$. Since the $a$-axis is the easy axis in NdRu$_2$Al$_{10}$, the metamagnetic transition is expected to be analogous to a spin-flop transition. In the field range of 2.6 T $< B_a <$ 4.6 T, AF-correlated moments align within the $bc$-plane, eliminating the sublattice-dependent effective field along the $a$-axis and resulting in a small $\gamma$. However, a finite $\gamma$, whose sign depends on the AF domain configuration, persists above $B_{sat}$ for $B \parallel a$ (Fig. 2(e)). Although the exact origin of this residual $\gamma$ remains unclear, it may stem from residual AF correlations at high fields. Eventually, under sufficiently strong fields, the AF correlations are fully suppressed, and thus $\gamma$ vanishes. A similar $\gamma$ behavior linked to AF correlations also appears in temperature-dependent data (Fig. 1(e)), developing slightly above $T_N$.

The above-mentioned results are reproducible across multiple devices [30] (See Supplemental Material at [URL will be inserted by publisher] for Reproducibility of nonreciprocal transport in other devices). Notably, all features of $\gamma$ observed in our experiments differ from those typically observed in magnetic-field-induced nonreciprocal charge transport; however, they are consistent with the expected behavior of a nonreciprocal response arising from AF order. Hence, our results clearly indicate that the observed $\gamma$ originates from the breaking of both TRS ($\mathcal{T}$) and SIS ($\mathcal{P}$) owing to the magnetic toroidal dipole induced by the AF order in NdRu$_2$Al$_{10}$. Although $\mathcal{T}$ is broken in this situation, the anomalous Hall effect at zero field is prohibited in typical magnetic toroidal dipole order. This is because the Berry curvature vanishes due to the preserved $\mathcal{PT}$ symmetry, which arises from the simultaneous breaking of both $\mathcal{T}$ and $\mathcal{P}$ [28].

In this material, additional phase transitions occur at lower temperatures (1.2 K and 0.9 K [33]). Although their nature is not yet fully clarified, these transitions may affect the nonreciprocal charge transport. Since its magnitude and sign are highly sensitive to the internal magnetic field $B_{\text{eff}}$, low-temperature nonreciprocal transport measurements can provide valuable insight into the underlying magnetic structures.

Finally, we compare the magnitude of the nonreciprocal charge transport observed herein with those of other reported materials. Figure 3(a) illustrates a plot of the $\gamma$ values of several materials, including NdRu$_2$Al$_{10}$. Except for the value of $\gamma$ on NdRu$_2$Al$_{10}$, the values of $\gamma$ at $B = 1$ T were used in this plot. Additionally, the maximum $\gamma$ value exhibited by each material was used for comparison, rather than the $\gamma$ value observed at the same temperature for all materials. A trend is evident; the material with a larger $\rho$ has a larger $\gamma$. This behavior can be understood based on the following expression, derived from the current density considering the second order of $E$: $j = \sigma_{(1)}E + \sigma_{(2)}E^2$, where $\sigma_{(1)}$ [$\Omega^{-1}$m$^{-1}$] denotes the normal electrical conductivity and $\sigma_{(2)}$ [$\Omega^{-2}$A$^{-1}$] represents the nonreciprocal electrical conductivity. Upon approximating $E \approx j/\sigma_{(1)}$, $E$ becomes [7]

$$E \approx \frac{1}{\sigma_{(1)}}j - \frac{\sigma_{(2)}}{\sigma_{(1)}^3}j^2 \quad (2)$$

$$= \rho j + (-\rho^3 \sigma_{(2)})j^2, \quad (3)$$

where $\sigma_{(1)} = 1/\rho$. A comparison between Eqs. (1) and (3) yields $|\gamma| = \rho^3|\sigma_{(2)}|$. As illustrated in Fig. 3(a), $\gamma$ are approximately proportional to $\rho^3$. For instance, in the case of Te, $\sigma_{(2)}$ obtained from two independent studies [11,16] are similar values of $\sigma_{(2)}$ on the order of approximately $10^{-4}$ $\Omega^{-2}$A$^{-1}$; however, their reported $\gamma$ (or $\gamma$') values varied by several orders [11,16]. This suggests that the coefficient indicating the magnitude of the intrinsic nonreciprocal charge transport is not $\gamma$, but rather $\sigma_{(2)}$. Meanwhile, materials with magnetic order—such as MnSi and the magnetically ordered state of CrNb$_3$S$_6$ (CNS$_{\text{mag}}$)—exhibit substantially larger $\sigma_{(2)}$ values compared to nonmagnetic materials. This implies that the effective internal field created by magnetic order enhances nonreciprocal charge transport [8,15]. For the nonmagnetic material ZrTe$_5$, $\sigma_{(2)}$ is reported to be comparable to that of magnetic materials owing to the topological semimetallic state with an extremely low carrier density [17].

Compared to the field-induced cases, we observed that NdRu$_2$Al$_{10}$ exhibits a much higher value of $\sigma_{(2)}$ on the order of approximately $10^3$–$10^4$ $\Omega^{-2}$A$^{-1}$ even in the absence of a magnetic field (Table 1). In the case of field-induced nonreciprocal transport, $\sigma_{(2)}$ has been derived within the framework of the Boltzmann equation and generally depends on the details of the band parameters, including the strength of spin–orbit coupling ($\lambda$) and the effective mass ($m^*$). Previous studies [7,13] have shown that $\sigma_{(2)}$ approximately scales with $\lambda$ and $m^*$. For NdRu$_2$Al$_{10}$, such band parameters have not yet been clarified; however, the effective mass of the isostructural compound NdRu$_2$Al$_{10}$ [34,35] may serve as a reasonable reference, since the 4$f$ electrons of Nd are well localized [24] and the conduction is mainly governed by itinerant Ru 4$d$ and Al 3$p$ electrons, as in the reference compound LaRu$_2$Al$_{10}$, which lacks 4$f$ electrons. Considering also the absence of heavy elements with strong spin–orbit coupling in NdRu$_2$Al$_{10}$, its band parameters are likely not particularly unusual. We therefore attribute the large $\sigma_{(2)}$ observed here to the strong effective magnetic field acting on the conduction bands through $c$–$f$ exchange interaction between the conduction electrons and the localized Nd moments.

Meanwhile, the emergence of $\gamma$ in terms of the magnetic structure of NdRu$_2$Al$_{10}$ remains nontrivial. Although the magnetic structure of NdRu$_2$Al$_{10}$ is yet to be elucidated, some sister compounds with the same crystal structure offer some insights. For instance, in NdFe$_2$Al$_{10}$ [36] and NdOs$_2$Al$_{10}$ [37], featuring the same crystal structure, the magnetic propagation vectors ($q_M = (q_a, q_b, q_c)$) are reported to be $q_{M1} = (0, 1/4, 0)$ and $q_{M2} = (0, 3/4, 0)$ for NdFe$_2$Al$_{10}$ [36] and $(0, 0.723, 0)$ for NdOs$_2$Al$_{10}$ [37]. In these cases, the total nonreciprocal response, which is the sum of individual $\gamma$ from each zigzag chain, is compensated by the modulation of $\gamma$ along the $b$-axis because the magnetic superstructure is either an even multiple or incommensurate with the original unit cell. If the $q_b$ parameter of NdRu$_2$Al$_{10}$ is an odd multiple, similar to the reported $q_M = (0, 4/5, 0)$ for the sister compound TbFe$_2$Al$_{10}$, the total compensation of $\gamma$ can be avoided [38]. This is also possible if the material contains an additional interchain modulation, such as a charge-density wave (CDW). In fact, previous study suggests the presence of a CDW transition in NdRu$_2$Al$_{10}$ [39], although the modulation wave number ($q_C$) is yet to be clarified. If we assume that these two modulation vectors are identical (i.e., $q_C = q_M$), the AF order would effectively exhibit $q_M = (0, 0, 0)$, fulfilling the conditions for the emergence of a ferroic toroidal dipole order and thus enabling nonreciprocal responses. Comprehensive (polarized) neutron and X-ray experiments are required to elucidate the details of the AF ordered state in this material, which could contribute to the quantitative understanding of nonreciprocal transport.

In summary, we observed nonreciprocal resistivity $\gamma$ in the AF phase of NdRu$_2$Al$_{10}$ in the absence of an external magnetic field. This zero-field nonreciprocal charge transport is attributed to the breaking of TRS and SIS owing to the AF order, which can be interpreted as a magnetic toroidal dipole order from the perspective of multipole. Moreover, the observed nonreciprocal conductivity $\sigma_{(2)}$ of NdRu$_2$Al$_{10}$ is considerably larger than previously reported values under external magnetic field, attributed to the strong effective magnetic field generated via $c$–$f$ exchange interaction. Furthermore, our results indicate that the nonreciprocal resistivity is governed by the spin configuration of AF domains. This result suggests that nonreciprocal resistivity can be utilized for the electrical readout of AF domains, a key technique for advancing AF spintronics [40–42].

**Acknowledgements**

We would like to thank K. Ogushi, Y. Onose, N. Kimura, H. Nojiri, and H. Kusunose for useful discussions. This work was partly supported by JSPS and MEXT KAKENHI with project Nos. JP25H00599, JP24K21522, JP23H04014, JP23H04868, JP23KK0052, JP23K22447, JP22H04933, JP22H01176, JP21H05470 and JP23K20824 and the GP-Spin program of Tohoku University. Moreover, the work was performed under the GIMRT Program of the Institute for Materials Research, Tohoku University (proposal Nos. 202212-HMKPA-0418, 202112-HMCPA-0408, and 202012-HMKPA-0417). KS was supported in part by the JSPS Research Fellowship Program (project No. 22J11892).

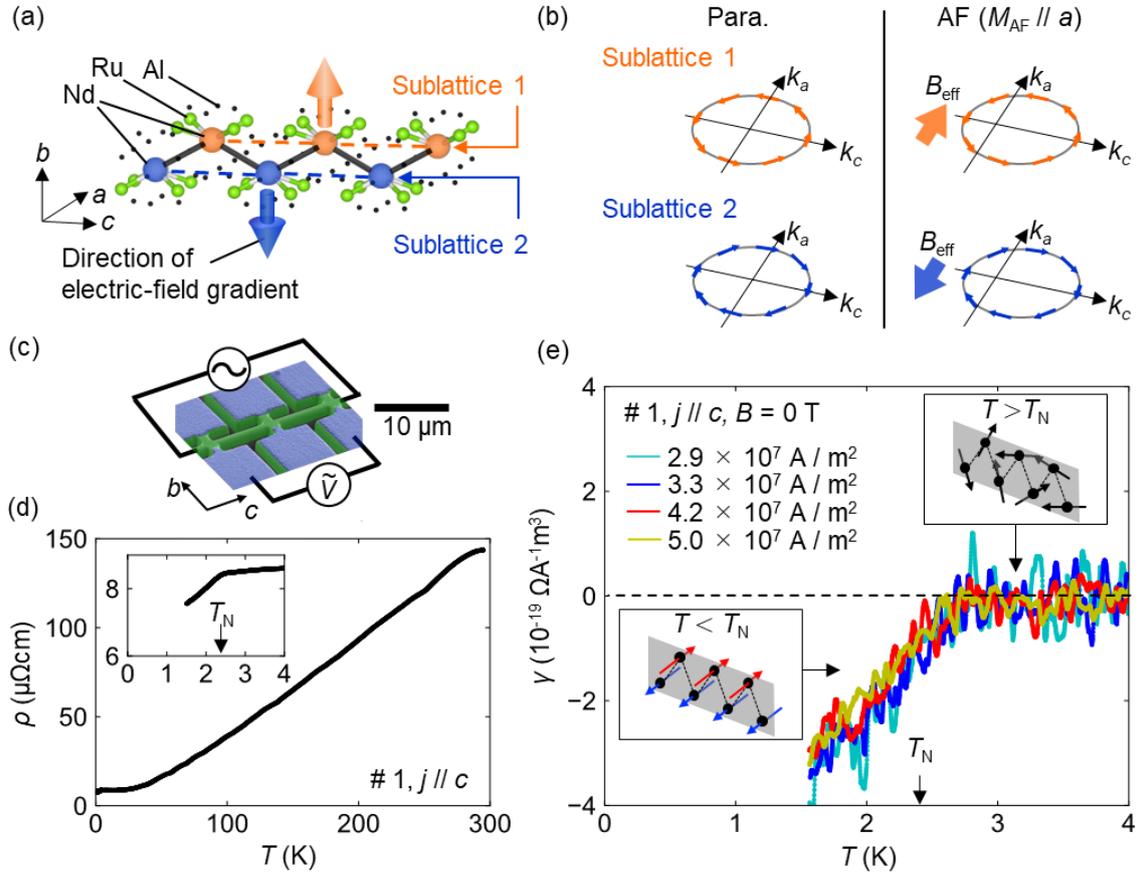

Figure 1: (a) Zigzag structure of NdRu$_2$Al$_{10}$. Nd atoms (indicated in orange and blue), each coordinated by four Ru atoms (indicated in light green), form a zigzag chain parallel to the *c*-axis. Sublattices 1 and 2 are constructed by Nd atoms presented in orange and blue, respectively. Bold orange and blue arrows indicate the sublattice-dependent directions of electric-field gradient, parallel to +*b* ([010]) and −*b* ([0−10]), respectively. (b) Schematic of the spin-polarized FS for each sublattice band. Due to the electric field gradient along the ±*b*-axis, Rashba type spin polarization is induced in both paramagnetic and AF state. (c) SEM image of the microfabricated device. Green and blue parts indicate NdRu$_2$Al$_{10}$ and sputtered gold, respectively. (d) Temperature dependence of electrical resistivity $\rho$ from 1.5 to 300 K, measured for device #1. The inset shows the change in resistivity associated with the AF transition. (e) Temperature dependence of nonreciprocal resistivity $\gamma$ from 1.5 to 4 K. Each temperature-dependent dataset for different current densities was obtained from separate temperature sweeps across $T_N$, confirming that the value of $\gamma$ at a given device position is well reproduced irrespective of the thermal history. The optimum current density for the measurement was chosen such that the Joule heating effect on $\gamma$ is negligible.

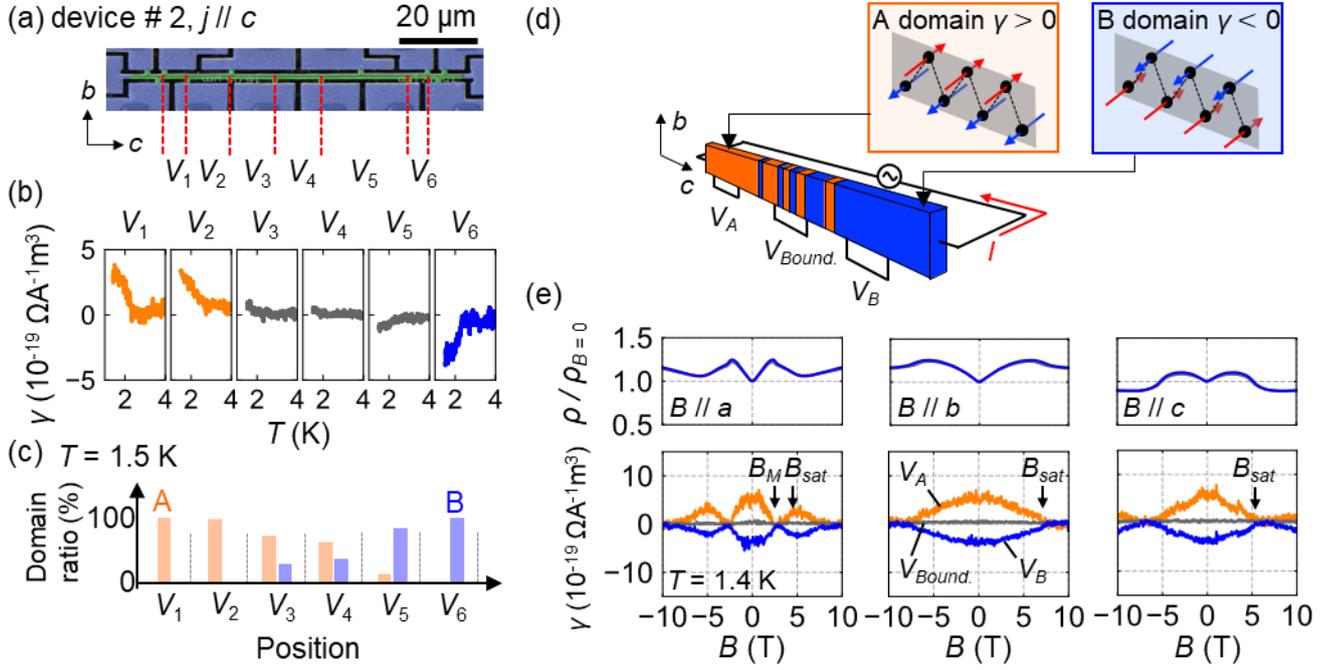

Figure 2: (a) SEM image of device #2. This design enables measurements of the position dependence of voltage signals using multiple voltage-contact pairs. (b) Temperature dependence of $\gamma$ at each voltage-contact pair ($V_1$–$V_6$) is indicated in (a). (c) Estimated AF domain ratio between domains A (↑↓↑↓) and B (↓↑↓↑), based on the results illustrated in Fig. 2(b). The domain ratio was calculated assuming that the signals from $V_1$ and $V_6$ correspond to 100% A and B domains, respectively. (d) A schematic illustration of AF domain distribution, which explains the location dependence of $\gamma$ shown in (b). Notably, the $\gamma$ values obtained from $V_A$ and $V_B$ exhibit opposite signs but equal magnitudes, while $V_{Bound.}$ shows a small $\gamma$ value owing to compensation resulting from domain mixing. (e) Magnetic-field dependence of $\rho/\rho_{B=0}$ (upper panels) and $\gamma$ (lower panels).

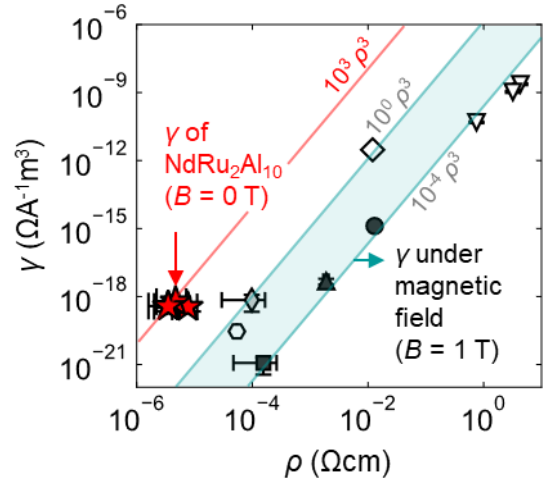

Figure 3: (a) Correlation plot between normal resistivity $\rho$ and nonreciprocal resistivity $\gamma$. For $\gamma$ values under an external magnetic field, the values at 1 T were used for comparison. The $\gamma$ values of Te, BiTeBr, MnSi, CrNb$_3$S$_6$, and ZrTe$_5$ are taken from the literature [7,8,11,15–17]. The range of $\sigma_{(2)}$ values (from $10^{-4}$ $\Omega^{-2}$A$^{-1}$ to $10^0$ $\Omega^{-2}$A$^{-1}$) observed in previous studies is indicated by shade. Red solid line indicates $\sigma_{(2)}$ value of this work ($10^3$ $\Omega^{-2}$A$^{-1}$).

Table 1: The comparison of nonreciprocal conductivity $\sigma_{(2)}$ between NdRu$_2$Al$_{10}$ (this work) and the other conductive materials (previous reports). The value of $\sigma_{(2)}$ for NdRu$_2$Al$_{10}$ is obtained from the average of four distinct devices (#1–#4), with the associated errors corresponding to their standard errors.

| | Material | $\sigma_{(2)}$ [$\Omega^{-2}$A$^{-1}$] | $T$ [K] | Ref. |
|---|---|---|---|---|
| ★ | NdRu$_2$Al$_{10}$ | $(6.0 \pm 1.9) \times 10^3$ | 1.5-1.6 | This work |
| ◇ | ZrTe$_5$ | $1.7 \times 10^0$ | 3 | [17] |
| ◆ | CNS $_{mag}$ | $8.8 \times 10^{-1}$ | 110 | [15] |
| ⬡ | MnSi | $1.8 \times 10^{-1}$ | 35 | [8] |
| ▲ | BiTeBr | $7.0 \times 10^{-4}$ | 2 | [7] |
| ● | Te (NW) | $5.9 \times 10^{-4}$ | 10 | [11] |
| ■ | CNS $_{para}$ | $3.2 \times 10^{-4}$ | 200 | [15] |
| ▽ | Te | $1.2 \times 10^{-4}$ | 2 | [16] |